\documentclass[journal]{IEEEtran}

\usepackage{graphicx}
\usepackage{amsmath, amssymb, amstext, verbatim, amsopn, cite, subfigure, lipsum}
\usepackage{balance}
\usepackage{url}
\usepackage{amsfonts}
\usepackage{epsfig}

\usepackage{algorithmic}
\usepackage{algorithm}

\usepackage{setspace}
\usepackage{stmaryrd}
\usepackage{psfrag}	
\usepackage{multirow}
\usepackage{multicol}
\usepackage{float}
\usepackage[process=auto]{pstool}
\usepackage{etoolbox}
\usepackage{color}
\usepackage{booktabs}
\usepackage{tabularx}

\allowdisplaybreaks

\usepackage{tikz}
\usetikzlibrary{arrows, decorations.markings}

\allowdisplaybreaks
\graphicspath{{images/}}

\newtheorem{theorem}{Theorem}

\newtheorem{remark}{Remark}

\newcommand*{\norm}[1]{\mathopen\| #1 \mathclose\|}
\newcommand*{\abs}[1]{\mathopen| #1 \mathclose|}

\def\norm#1{\mathopen\| #1 \mathclose\|}

\newcommand{\vv}[1]{\boldsymbol{\mathrm{#1}}}
\newcommand{\mm}[1]{\boldsymbol{\mathrm{#1}}}

\newcommand{\expect}[1]{{\mathbb{E}}\!\left\{ #1 \right\}}

\newcommand{\herm}{{\sf{H}}}
\newcommand{\transp}{{\sf{T}}}
\ifCLASSINFOpdf
\else
\fi
\begin{document}
%
\title{Max-Min Multi-Cell Aware Precoding and Power Allocation for Downlink Massive MIMO Systems \vspace*{0mm}}
%
%
%
\author{Shahram Zarei, Jocelyn Aulin, Wolfgang Gerstacker, and Robert Schober \vspace*{-6mm} 
\thanks{S. Zarei, W. Gerstacker, and R. Schober are with Friedrich-Alexander-University Erlangen-N{\"u}rnberg, Erlangen, Germany (e-mail: shahram.zarei@fau.de; wolfgang.gerstacker@fau.de; robert.schober@fau.de).}
\thanks{J. Aulin is with Huawei Technologies Sweden AB, Gothenburg, Sweden (e-mail: jocelyn.aulin@huawei.com).}
}
\maketitle
\begin{abstract}
We propose a max-min multi-cell aware regularized zero-forcing (Max-Min MCA-RZF) precoding and power allocation scheme for downlink multi-cell massive multiple-input multiple-output (MIMO) systems. A general correlated channel model is considered, and the adopted channel state information (CSI) acquisition model includes the effects of estimation errors and pilot contamination. We use results from random matrix theory to derive deterministic equivalents for the proposed Max-Min power allocation in the large system limit which solely depend on statistical CSI, but not on individual channel realizations. Our numerical results show that the proposed Max-Min MCA-RZF precoder achieves a substantially higher network-wide minimum rate than the MCA-RZF and the conventional RZF precoders with uniform power allocation, respectively, as well as the conventional RZF precoder with Max-Min power allocation. 
\end{abstract}
%
\IEEEpeerreviewmaketitle
%
\vspace*{-5mm}
\section{Introduction}\label{Sec_Intro}
\IEEEPARstart{F}{airness} is a very important aspect when designing wireless networks and in general refers to providing the same quality of service (QoS) to all user terminals (UTs). Fairness can be enforced by maximizing the minimum achievable rate (Max-Min) among the users. Moreover, in downlink (DL) massive multiuser multiple-input multiple-output (MU-MIMO) systems, linear precoders are attractive choices since they provide a good compromise between complexity and performance. Regularized zero-forcing (RZF) precoding is a widely used linear precoding technique which offers a good performance in single-cell scenarios where the number of UTs is much smaller than the number of base station (BS) antennas \cite{Wagner2012, Hoydis2013}. However, in multi-cell scenarios, the performance of the RZF precoder can be severely degraded by multi-cell interference and pilot contamination \cite{Marzetta2010}. The above considerations motivate us to design a linear Max-Min precoder which is aware of multi-cell interference. 

One of the pioneering works on multi-cell precoding for MU-MIMO systems is \cite{Dahrouj_TWC2010}. Here, the authors propose a coordinated precoding scheme which minimizes the weighted transmit power under target signal-to-interference-plus-noise ratio (SINR) constraints. The authors in \cite{Bjoernson_SigProcMag2014} provide a general structure for the optimal linear precoding vectors as a function of certain Lagrange multipliers, which have to be optimized according to the adopted optimization criterion. Moreover, in \cite{Zakhour2012}, maximization of the network-wide minimum achievable rate for a \emph{two-cell} network is studied. In \cite{Lakshminarayana_TIT2015}, \cite{Sanguinetti_TWC2016}, a total transmit power minimizing multi-cell precoder for given target rates is proposed.
In \cite{Tan_ACM2013}, several power allocation schemes including an efficient Max-Min power allocation algorithm are proposed for general wireless networks composed of a number of wireless links with given gains. 
Another relevant work is \cite{Huang_TWC2013}, where the authors consider joint beamforming and Max-Min power control for DL MU-MIMO systems. A recent work on power control for multi-cell massive MIMO systems is \cite{Guo_ICC2014}, where the authors propose an iterative power allocation and detection scheme which minimizes the sum transmit power of all UTs subject to per-UT SINR and power constraints in an uplink (UL) multi-cell massive MIMO system. In \cite{Zhang_TWC2015}, the authors present a sum rate maximizing power allocation scheme for UL and DL massive MIMO systems. However, multi-cell aware precoding with Max-Min power allocation in case of \emph{correlated} channels and \emph{imperfect} CSI knowledge, and specifically, in the presence of pilot contamination, has not been investigated, yet. In this letter, we propose a multi-cell interference aware RZF precoder with Max-Min power allocation (Max-Min MCA-RZF) which maximizes the network-wide minimum achievable rate.
Taking into account the large numbers of BS antennas which are typical for massive MIMO systems, we apply results from random matrix theory and derive asymptotic expressions for the Max-Min power allocation in the large system limit. The derived asymptotic power allocation depends on statistical CSI, but not on individual channel realizations. Hence, it has to be updated less frequently than the channel estimates.

In contrast to \cite{Dahrouj_TWC2010}, where a weighted sum transmit power minimizing precoder for \emph{perfect} CSI is proposed, in this work, we develop a minimum rate maximizing and multi-cell interference aware precoder, where the power allocation depends merely on the statistical CSI and not on the individual channel realizations. Different from \cite{Bjoernson_SigProcMag2014}, where a general structure for optimal linear precoding vectors is proposed, which requires \emph{perfect} CSI and includes unknown Lagrangian multipliers, we derive a Max-Min multi-cell aware precoder and the corresponding power allocation for \emph{imperfect} CSI. In contrast to \cite{Zakhour2012}, which considers a two-cell network with \emph{uncorrelated} channels and \emph{perfect} CSI knowledge at the BSs, we consider a more general system model with an arbitrary number of cells, \emph{correlated} channels, and \emph{imperfect} CSI. Moreover, as opposed to \cite{Lakshminarayana_TIT2015}, \cite{Sanguinetti_TWC2016}, where the authors minimize the total transmit power for \emph{uncorrelated} channels and replace the actual channel vectors with the estimated ones in the expression for the derived precoding vectors to deal with \emph{imperfect} CSI, we derive the optimal minimum rate maximizing precoding vectors for \emph{correlated} channels, which account for multi-cell interference and \emph{imperfect} CSI. Contrary to \cite{Tan_ACM2013}, where a Max-Min power allocation scheme for a general network consisting of wireless links characterized only by a link gain is proposed, we develop both multi-cell interference aware precoding and Max-Min power allocation for a practical multi-cell massive MIMO scenario with \emph{imperfect} CSI, and derive deterministic expressions for the power allocation in the large system limit. Moreover, as opposed to \cite{Huang_TWC2013}, where a system model with \emph{perfect} CSI and \emph{uncorrelated} channels is considered, we adopt a correlated channel model and pilot contamination. Furthermore, in contrast to \cite{Guo_ICC2014}, where the authors minimize the total sum transmit power subject to target SINRs in an UL multi-cell massive MIMO system, we propose a Max-Min MCA-RZF precoder which maximizes the minimum achievable rate in the DL of a multi-cell massive MIMO system. Different from \cite{Zhang_TWC2015}, where the authors propose a sum rate maximizing power control for the UL and DL of massive MIMO systems, our proposed Max-Min MCA-RZF scheme maximizes the network-wide minimum achievable rate.

\emph{Notation:} Boldface lower and upper case letters represent column vectors and matrices, respectively. $\mm{I}_K$ denotes the $K \times K$ identity matrix and ${\left[\mm{A}\right]}_{k,:}$, ${\left[\mm{A}\right]}_{:,l}$, and ${\left[\mm{A}\right]}_{k, l}$ stand for the $k$th row, the $l$th column, and the element in the $k$th row and the $l$th column of matrix $\mm{A}$, respectively. $(\cdot)^*$ denotes the complex conjugate, and  $\mathrm{tr}(\cdot)$ and $(\cdot)^{\herm}$ represent the trace and Hermitian transpose of a matrix, respectively. $\expect{\cdot}$ stands for the expectation operator and $\mathcal{C} \mathcal{N} \left(\vv{u}, \mm{\Phi} \right)$ denotes a circular symmetric complex Gaussian distribution with mean vector $\vv{u}$ and covariance matrix $\mm{\Phi}$. Moreover, ``a.s." stands for ``almost sure" convergence and $a \asymp b$ indicates $a-b \xrightarrow[N\rightarrow\infty]{\mathrm{a.s.}} 0$.
%
\vspace*{-2mm}
\section{System Model}\label{Sec_SysModelBench}
In this letter, we consider the DL of a time-division duplex (TDD) multi-cell massive MIMO system with universal frequency reuse. The number of cells is denoted by $L$, and in each cell, a BS equipped with $N$ antennas simultaneously transmits data to $K$ single-antenna UTs. The independent and identically distributed (i.i.d.) zero-mean complex Gaussian data symbols intended for transmission to the UTs in the $l$th cell are stacked into vector $\vv{d}_l  = [d_{l1}, \ldots ,d_{lK}]^\transp$ with $\mathbb{E} \left\lbrace \vv{d}_l \vv{d}_l^\herm \right\rbrace= \mm{I}_K$, where $d_{lk}$ is the data symbol of the $k$th UT in the $l$th cell. The vector of the stacked received data symbols of the UTs in the $j$th cell is given by
\begin{align}
{\hat{\vv{d}}_{\mathrm{DL}, j}} =  \mathop{ \sum_{l=1}^{L}} \mm{G}^\herm_{lj} \mm{V}_{l} \mm{P}_l^{1/2} \vv{d}_{l} + \vv{z}_j,
\label{Eqn_DL_Sig_RX} 
\end{align}
where $\mm{G}_{lj} \hspace*{-1mm} = \hspace*{-1mm} \left[\vv{g}_{lj1}, \ldots, \vv{g}_{ljK}\right] \in \mathbb{C}^{N \times K} $ and $\mm{V}_l \hspace*{-1mm} = \hspace*{-1mm} \left[ \vv{v}_{l1}, \ldots, \vv{v}_{lK} \right] \in \mathbb{C}^{N \times K}$ denote the channel matrix between the UTs in the $j$th cell and the $l$th BS and the precoding matrix at the $l$th BS, respectively, with $\vv{v}_{lk}$ being the unit-norm precoding vector for the $k$th UT at the $l$th BS. In (\ref{Eqn_DL_Sig_RX}), $\mm{P}_l=\mathrm{diag} \left(p_{l1},\ldots,p_{lK}\right)$ represents the DL power allocation matrix, where $p_{lk}$ is the power allocated to the $k$th UT in the $l$th cell. Here, we consider a total transmit power constraint $\sum_{l=1}^L\sum_{k=1}^K p_{lk} \leq LK\rho_\mathrm{DL}$, where we define $\rho_\mathrm{DL}$ to be the average per-user transmit signal-to-noise ratio (SNR). Moreover, $\vv{z}_j \sim \mathcal{C}\mathcal{N}\left(\vv{0}, \mm{I}_K\right)$ represents the stacked vector of the additive white Gaussian noise (AWGN) samples at the UTs in the $j$th cell. In this work, we assume a block flat fading channel and a correlated channel model, i.e., $\vv{g}_{ljk} = \tilde{\mm{R}}_{ljk} \vv{h}_{ljk}$, where $\vv{h}_{ljk} \sim \mathcal{C} \mathcal{N} \left(\mm{0}, \mm{I}_N \right) $, and $\mm{R}_{ljk} = \tilde{\mm{R}}_{ljk} \tilde{\mm{R}}_{ljk}^\herm = \mathbb{E} \lbrace \vv{g}_{ljk} \vv{g}^\herm_{ljk} \rbrace$ represents the covariance matrix of the channel between the $k$th UT in the $j$th cell and the $l$th BS. 
Assuming linear minimum mean square error (LMMSE) channel estimation and considering the effect of pilot contamination, the estimated channel vector between the $k$th UT in the $j$th cell and the $l$th BS is given by \cite{Kay_StSiProDet2013, Hoydis2013}
\begin{align}
\hat{\vv{g}}_{ljk} = \mm{\Omega}_{ljk} \left( \sum_{m=1}^{L} \vv{g}_{lmk} + \frac{1}{\sqrt{\rho_\mathrm{TR}}} \check{\vv{z}}_{lk} \right),
\label{Eqn_g_hat}
\end{align}
where $\mm{\Omega}_{ljk}$ is defined as
\begin{align}
\mm{\Omega}_{ljk} \triangleq \mm{R}_{ljk} \left( \sum_{m=1}^{L} \mm{R}_{lmk} + \frac{1}{\rho_\mathrm{TR}} \mm{I}_N \right)^{-1},
\label{Eqn_Omega}
\end{align}
with $\rho_\mathrm{TR}$ and $\check{\vv{z}}_{lk} \hspace*{-1mm} \sim \hspace*{-1mm} \mathcal{C} \mathcal{N} \hspace*{-1mm} \left( \vv{0}, \mm{I}_N \right)$ being the training SNR and the channel estimation noise at the $l$th BS corresponding to the $k$th UT, respectively. 
The performance metric used in this letter is the network-wide minimum achievable rate which is defined as 
\begin{align}
R \triangleq \min_{j \in \left\lbrace 1,\ldots, L \right\rbrace, k \in \left\lbrace 1,\ldots, K \right\rbrace } R_{jk} = \log_2 \left( 1 + \mathrm{SINR}_{\mathrm{DL}, jk} \right),
\label{Eqn_MinRate_Ergo}
\end{align}
where the SINR at the $k$th UT in the $j$th cell in the DL is defined as \cite{Jose_TWC2011}
\begin{flalign}
&\mathrm{SINR}_{\mathrm{DL}, jk} \triangleq& \nonumber
\end{flalign}
\begin{equation}
\frac{ p_{jk} \Big| \mathbb{E}_{\mm{G}} \left\lbrace {\vv{g}_{jjk}^\herm \mm{v}_{jk}} \right\rbrace \Big| ^2}{ \displaystyle \mathop{ \sum_{l=1}^{L} \hspace*{-.7mm} \sum_{q=1}^{K}} \hspace*{-.5mm} p_{lq} \mathbb{E}_{\mm{G}} \hspace*{-.5mm} \left\lbrace \big| {\vv{g}_{ljk}^\herm \mm{v}_{lq}} \big| ^2 \right\rbrace \hspace*{-1mm}-\hspace*{-1mm} p_{jk} \Big| \mathbb{E}_{\mm{G}} \hspace*{-.5mm} \left\lbrace {\vv{g}_{jjk}^\herm \mm{v}_{jk}} \hspace*{-.3mm} \right\rbrace \Big| ^2 \hspace*{-1.5mm}+\hspace*{-1mm} 1}, \vspace*{-2mm}
\label{Eqn_SINR_DL}
\end{equation}
where $\mm{G}=\left[ \mm{G}_1 \cdots \mm{G}_L \right]$ and $\mm{G}_l=\left[\mm{G}_{l1} \cdots \mm{G}_{lL}\right]$.
%
\vspace*{-1mm}
\section{Max-Min MCA-RZF Precoder}\label{Sec_MCA_Prec}
In this section, we derive the proposed Max-Min MCA-RZF precoder. The optimization objective in this letter is the maximization of the network-wide minimum achievable rate. The corresponding optimization problem can be expressed as\footnote{The optimization problem in (\ref{Eqn_MaxMinOptProblem}) leads to equal achievable rates for all UTs which may compromise the achievable sum rate. Considering more advanced optimization criteria which enable a trade off between achievable sum rate and minimum rate is an interesting topic for future work.}
\begin{align}
& \max_{\mm{V}_{l}, \mm{P}_{l}, \forall l} \ \min_{j, k} \  \log_2 \left( 1+\mathrm{SINR}_{\mathrm{DL}, jk} \right) \nonumber \\ \vspace*{-3mm}
\mathrm{s.t.} \ \ & \displaystyle \mathop{ \sum_{l=1}^{L} \sum_{q=1}^{K}} p_{lq} \leq L K \rho_\mathrm{DL}, \nonumber \\
& \vv{v}_{lq}^\herm \vv{v}_{lq} =1, \ \forall l, q.
\label{Eqn_MaxMinOptProblem}
\end{align}
\subsection{Transformation of Optimization Problem}\label{Sec_TransOptProblem} \vspace*{-0mm}
The optimization problem in (\ref{Eqn_MaxMinOptProblem}) is difficult to solve due to the coupling of the UTs introduced by the precoding vectors. Hence, we apply the UL/DL duality framework presented in \cite[Theorem 3]{Bjoernson_TWC2016} to transform (\ref{Eqn_MaxMinOptProblem}) into its dual UL equivalent which is easier to solve. 
In the dual UL system, the channel matrix between the UTs in the $j$th cell and the $l$th BS and the detection matrix at the $l$th BS are given by $\mm{G}_{lj}$ and $\mm{V}^\herm_l$, respectively. In particular, if the vector containing the powers of all UTs in all cells in the DL, i.e., $\vv{p}=\left[ \vv{p}^\transp_1, \ldots, \vv{p}^\transp_L \right]^\transp$, where $\vv{p}_l =\left[p_{l1},\ldots,p_{lK}\right]^\transp, l \in \left\lbrace1,\ldots,L\right\rbrace$, is chosen as
\begin{align}
\vv{p} = \left( \mm{I}_{KL} - \mathrm{diag}\left( \vv{a} \right) \cdot \mm{A}^\transp \right)^{-1} \vv{a},
\label{Eqn_Vector_p}
\end{align}
for the same sum power\footnote{The power constraint in (\ref{Eqn_MaxMinOptProblem}) is a total power constraint over all BSs in all cells. If this constraint is changed to a per-BS power constraint, the UL/DL duality cannot be used anymore and the resulting optimization problem would become much more complicated.}, the same per-UT SINR can be achieved in the DL as in the UL. 
In (\ref{Eqn_Vector_p}), the elements of vector $\vv{a} = \left[ a_{1},\ldots,a_{KL} \right]^\transp$ are given by
\begin{align}
\left[\vv{a}\right]_{\left(j-1\right)K+k} \hspace*{-1mm} \triangleq \hspace*{-1mm} \frac{ \mathrm{SINR}_{{\mathrm{UL}}, jk} }{ \Big| \mathbb{E}_{\mm{G}} \left\lbrace {\mm{v}_{jk}^\herm \vv{g}_{jjk} } \right\rbrace \Big| ^2}, \label{Eqn_Vec_a} 
\end{align}
where $j\in  \left\lbrace 1,\ldots,L \right\rbrace, k \in  \left\lbrace 1,\ldots,K \right\rbrace$, and the elements of matrix $\mm{A} \in \mathbb{R}^{KL \times KL}$ in (\ref{Eqn_Vector_p}) are defined as 
\begin{flalign}
&\left[\mm{A}\right]_{\left(j-1\right)K+k, \left(l-1\right)K+q} \triangleq& \nonumber
\end{flalign}
\begin{eqnarray}
\begin{cases}
  \mathbb{E}_{\mm{G}} \hspace*{-1mm} \left\lbrace \big| {\mm{v}_{jk}^\herm \vv{g}_{jlq}} \big| ^2 \right\rbrace \hspace*{-1mm}-\hspace*{-1mm}\Big| \mathbb{E}_{\mm{G}} \hspace*{-.5mm} \left\lbrace {\mm{v}_{jk}^\herm \vv{g}_{jjk} } \right\rbrace \hspace*{-.7mm} \Big| ^2 \hspace*{2mm} \mathrm{if} \ (j, k )\hspace*{-1mm}=\hspace*{-1mm}(l, q), & \\
  \mathbb{E}_{\mm{G}} \hspace*{-1mm} \left\lbrace \big| {\mm{v}_{jk}^\herm \vv{g}_{jlq}} \big| ^2 \right\rbrace \hspace*{35mm} \mathrm{otherwise}, 
\label{Eqn_Mat_A}
\end{cases} 
\end{eqnarray}
where $l\in  \left\lbrace 1,\ldots,L \right\rbrace, q \in  \left\lbrace 1,\ldots,K \right\rbrace$, and the SINR of the $k$th UT in the $j$th cell in the dual UL system is given by 
\vspace*{-1.5mm}
\begin{flalign}
& \mathrm{SINR}_{{\mathrm{UL}}, jk} \triangleq& \nonumber 
\end{flalign}
\begin{align}
& \frac{ \check{p}_{jk} \Big| \mathbb{E}_{\mm{G}} \left\lbrace {\mm{v}_{jk}^\herm \vv{g}_{jjk} } \right\rbrace \Big| ^2}{ \displaystyle \mathop{ \sum_{l=1}^{L} \hspace*{-.7mm} \sum_{q=1}^{K}} \hspace*{-.5mm} \check{p}_{lq} \mathbb{E}_{\mm{G}} \hspace*{-.5mm} \left\lbrace \big| {\mm{v}_{jk}^\herm \vv{g}_{jlq}} \big| ^2  \right\rbrace \hspace*{-1mm}-\hspace*{-1mm} \check{p}_{jk} \Big| \mathbb{E}_{\mm{G}} \hspace*{-.5mm} \left\lbrace {\mm{v}_{jk}^\herm \vv{g}_{jjk} } \right\rbrace \hspace*{-.7mm} \Big| ^2 \hspace*{-1.5mm}+ 1}
\label{Eqn_SINR_UL1}
\end{align}
with ${\check{p}}_{lk}$ being the transmit power of the $k$th UT in the $l$th cell. After applying the UL/DL duality theorem, the dual UL equivalent of the optimization problem in (\ref{Eqn_MaxMinOptProblem}) is given by
\begin{align}
& \max_{\mm{V}_l, \check{\mm{P}}_l, \forall l} \min_{j, k}  \log_2 \left( 1 + \mathrm{SINR}_{{\mathrm{UL}}, jk} \right) \nonumber \\
\mathrm{s.t.} \ \ &\displaystyle \mathop{ \sum_{l=1}^{L} \sum_{q=1}^{K}} {\check{p}}_{lq} \leq L K \rho_\mathrm{DL}, \nonumber \\
& \vv{v}_{jk}^\herm \vv{v}_{jk} =1, \ \forall j, k,
\label{Eqn_MaxMinOptProblem_UL}
\end{align}
where $\check{\mm{P}}_l=\mathrm{diag}\left( \check{p}_{l1},\ldots,\check{{p}}_{lK} \right)$ and $\mathrm{SINR}_{{\mathrm{UL}}, jk}$ is given by (\ref{Eqn_SINR_UL1}). Now, we derive the optimal detector vectors and transmit powers for the dual UL system.
\subsection{Max-Min MCA-RZF Detector in the Dual UL}\label{Sec_MCA_Det_UL} \vspace*{-2mm}
As can be seen from (\ref{Eqn_SINR_UL1}), in contrast to the DL, the detector vectors in the UL can be optimized individually which makes the UL problem easier to solve. Nevertheless, the joint optimization of the detector vectors $\vv{v}_{jk}$ and transmit powers ${\check{p}}_{lq}$ in (\ref{Eqn_MaxMinOptProblem_UL}) is still difficult, since the optimal detector vectors depend on the transmit powers and vice versa.
Here, in a first step, we find optimal detector vectors $\vv{v}_{jk}$ for given powers $\check{p}_{lq}$. Then, in a second step, using the detector vectors $\vv{v}_{jk}$ obtained in the first step, (\ref{Eqn_MaxMinOptProblem_UL}) is solved for the optimal powers $\check{p}_{lq}$. 
In the following theorem, we present the optimal detector vectors for a given power allocation in the dual UL system.
\vspace*{0mm}
\begin{theorem} For given LMMSE channel estimates $\hat{\vv{g}}_{jlq}$ and powers $\check{p}_{lq}, l \in \left\lbrace 1,\ldots,L\right\rbrace, q \in \left\lbrace 1,\ldots,K\right\rbrace,$ and in the large system limit, i.e. for $N \rightarrow \infty$, the solution of the optimization problem in (\ref{Eqn_MaxMinOptProblem_UL}) can be expressed as $\vv{v}_{jk}=\tilde{\vv{u}}_{jk} / \norm{\tilde{\vv{u}}_{jk}}$, where $\tilde{\vv{u}}_{jk}$ is given by
\begin{align}
\tilde{\vv{u}}_{jk} \hspace*{-.5mm}=\hspace*{-.5mm} \left( \sum_{l=1}^{L} \sum_{q=1}^K \check{p}_{lq} \left( {\hat{\mm{g}}}_{jlq} {\hat{\mm{g}}}^\herm_{jlq} + {\mm{\Delta}}_{jlq} \right) + \mm{I}_N  \right)^{-1} {\hat{\vv{g}}}_{jjk} ,\label{Eqn_Detector_PowAlloc}
\end{align}
with ${\mm{\Delta}}_{jlq}= \mm{R}_{jlq}-\mm{\Psi}_{jlq}$ being the covariance matrix of the estimation error for the channel between the $q$th UT in the $l$th cell and the $j$th BS. Here, $\mm{\Psi}_{jlq}$ is given by
\begin{equation}
\mm{\Psi}_{jlq} \hspace*{-0.5mm} \triangleq \mm{R}_{jlq}\left( \hspace*{-0.5mm} \sum_{m=1}^L \mm{R}_{jmq} \hspace*{-0.5mm} \hspace*{-0.5mm}+\hspace*{-0.5mm} \frac{1}{\rho_\mathrm{TR}} \mm{I}_N \right)^{-1} \hspace*{-1mm} \mm{R}_{jlq}.
\label{Eqn_Delta}
\end{equation}
\end{theorem}
\begin{IEEEproof}
Please refer to Appendix A. 
\end{IEEEproof}
Applying the detector vectors obtained in the previous step, the optimization problem in (\ref{Eqn_MaxMinOptProblem_UL}) becomes a function of $\check{p}_{lq}$, only, and can be solved by using fixed-point methods. Here, we adopt an efficient algorithm introduced in \cite{Tan_ACM2013} which has been proposed for general wireless networks and solves the convex dual of the optimization problem in (\ref{Eqn_MaxMinOptProblem_UL}). The algorithm in \cite{Tan_ACM2013} solves (\ref{Eqn_MaxMinOptProblem_UL}) for per-UT power constraints, i.e., $\check{p}_{lq} \leq \rho_\mathrm{DL} , \forall l, q,$ instead of the sum power constraint. This change in constraints restricts the feasible set of the optimization problem in (\ref{Eqn_MaxMinOptProblem_UL}), but is advocated here due to the excellent performance and low complexity of the resulting power allocation algorithm, cf. Section \ref{Sec_NumResults}. In particular, the optimal powers are obtained with the following iterative process
\begin{align}
\check{p}^{(i+1)}_{lq} = \frac{\check{p}^{(i)}_{lq}}{\mathrm{SINR}^{(i)}_{\mathrm{UL}, lq}}, \ \ \  \
\check{p}^{(i+1)}_{lq} \leftarrow \frac{\rho_\mathrm{DL} \check{p}^{(i+1)}_{lq}}{ \max_{l^\prime, q^\prime} \ \check{p}^{(i+1)}_{l^\prime q^\prime}}, \label{MaxMin_PowerConst}
\end{align}
where $i$ is the iteration index, and $\mathrm{SINR}^{(i)}_{\mathrm{UL}, lq}$ is obtained by substituting $\check{p}_{lq}=\check{p}^{(i)}_{lq}$ into (\ref{Eqn_SINR_UL1}). Note that the second step in (\ref{MaxMin_PowerConst}) is performed to ensure that the per-UT power constraint $\check{p}^{(i+1)}_{lq} \leq \rho_\mathrm{DL}$ is met.
\begin{remark} According to \cite[Theorems 2, 4]{Tan_ACM2013}, the above algorithm converges geometrically fast to the globally optimal solution for any initial value of the power allocation.
\end{remark}
The power allocation expression in (\ref{MaxMin_PowerConst}) is still difficult to compute due to the expectations in (\ref{Eqn_SINR_UL1}). Hence, in order to simplify the SINR calculation, in the following theorem, we provide a closed-form expression for the asymptotic value of $\mathrm{SINR}^{(i)}_{\mathrm{UL}, lq}$ in the large system limit, i.e., for $N \rightarrow \infty$ which we denote by $\mathrm{SINR}^{(i)^\circ}_{\mathrm{UL}, lq}$. The transmit powers $\check{p}_{lq}$ are then calculated by (\ref{MaxMin_PowerConst}) after replacing $\mathrm{SINR}^{(i)}_{\mathrm{UL}, lq}$ with $\mathrm{SINR}^{(i)^\circ}_{\mathrm{UL}, lq}$ in (\ref{MaxMin_PowerConst}). 
\vspace*{-0mm}
\begin{theorem}\label{Theorem_Asy_SINR}
For the dual UL system defined at the beginning of this section and in the large system limit, i.e., for $N \rightarrow \infty$, the asymptotic SINR of the $q$th UT in the $l$th cell in the $i$th iteration step is given by
\begin{align}
& \mathrm{SINR}^{(i)^\circ}_{\mathrm{UL}, lq} = \check{p}^{(i)}_{lq} \left( \check{\delta}^{(i)}_{lq} \right)^2 \Bigg/ \Bigg( \displaystyle{\sum_{m=1, m \neq l}^L }  \check{p}^{(i)}_{mq} \abs{\theta^{(i)}_{lmq}}^2 \nonumber \\
& +\hspace*{-1.5mm} \displaystyle{\sum_{m=1}^L \sum_{w=1, w \neq q}^K } \hspace*{-2mm} \check{p}^{(i)}_{m w} \Bigg( \hspace*{-.7mm} \zeta^{(i)}_{lm w} \hspace*{-.7mm}+\hspace*{-.7mm} \frac{\abs{\theta^{(i)}_{lm  w}}^2 \eta^{(i)}_{lw}}{\left(1\hspace*{-.7mm}+\hspace*{-.7mm}\delta^{(i)}_{ll w}\right)^2} \hspace*{-.7mm}-\hspace*{-.7mm} 2 \Re{\left\lbrace \hspace*{-.7mm} \frac{ (\theta^{(i)}_{lm w})^* \mu^{(i)}_{lm w} }{1\hspace*{-.7mm}+\hspace*{-.7mm}\delta^{(i)}_{ll w}} \hspace*{-.7mm} \right\rbrace} \hspace*{-.7mm} \Bigg) \nonumber \\
& +\hspace*{-1.0mm} { \bar{ \xi}_{lq}^{(i)}} \left( 1 \hspace*{-.7mm}+\hspace*{-.7mm} \delta^{(i)}_{llq} \right)^2 \hspace*{-.5mm} \Bigg),
\label{Eqn_Asy_SINR}
\end{align}
where $\delta_{llq}^{\left(i\right)}$ is the solution of the following set of fixed-point equations for $m=l$ and $w=q$
\begin{align}
& \delta^{(i)}_{lmw} \hspace*{-.5mm}=\hspace*{-.5mm} \frac{1}{N} \mathrm{tr} \left( \check{p}^{(i)}_{mw} \mm{\Psi}_{lmw} \mm{T}^{(i)}_l \right), \label{Eqn_deltak} \\ \vspace*{-2mm}
& \mm{T}^{(i)}_l \hspace*{-1mm}=\hspace*{-1.4mm} \left( \hspace*{-.5mm} \sum_{m=1}^L \hspace*{-1mm} \sum_{w=1}^{K} \hspace*{-.7mm} \check{p}^{(i)}_{mw} \hspace*{-1.3mm} \left( \hspace*{-.5mm} \frac{\mm{\Psi}_{lmw}}{N\left(1+\delta^{(i)}_{lmw} \hspace*{-.5mm} \right)}  \hspace*{-.5mm}+\hspace*{-.5mm} \mm{\Delta}_{lmw} \hspace*{-1mm} \right) \hspace*{-1.5mm}+\hspace*{-.7mm} \frac{\mm{I}_N}{N}  \hspace*{-.5mm} \right)^{-1}, \label{Eqn_T}
\end{align}
In (\ref{Eqn_Asy_SINR}), $\check{\delta}^{(i)}_{lq}$ and  $\theta^{(i)}_{lmw}$ are given by $\mathrm{tr} \left( \mm{B} \mm{T}^{(i)}_l \right)/N$ after replacing $\mm{B}$ with $\mm{\Psi}_{llq}$ and $\mm{R}_{lmw} \mm{\Omega}_{llw}$, respectively, where, $\mm{\Psi}_{llq}$ and $\mm{\Omega}_{llw}$ are given by (\ref{Eqn_Delta}) and (\ref{Eqn_Omega}), respectively. Furthermore, in (\ref{Eqn_Asy_SINR}), $ \zeta^{(i)}_{lmw}$, $\eta^{(i)}_{lw}$, and $\mu^{(i)}_{lmw}$ are given by $\mathrm{tr} \left( \mm{C} \check{\mm{T}}^{(i)}_{lq} \right)/N^2$ for $\mm{C}$ equal to $\mm{R}_{lmw}$, $\check{p}^{(i)}_{lw} \mm{\Psi}_{llw}$, and $\check{p}^{(i)}_{lw} \mm{\Omega}_{lmw} \mm{R}_{llw} $, respectively, and $\check{\mm{T}}^{(i)}_{lq}$ is computed as
\begin{align}
\hspace*{-2mm} \check{\mm{T}}^{(i)}_{lq} \hspace*{-1mm}=\hspace*{-1mm} \mm{T}^{(i)}_l \mm{\Psi}_{llq} \mm{T}^{(i)}_l \hspace*{-1mm}+\hspace*{-2mm} \sum_{m=1}^L \hspace*{-.5mm} \sum_{w=1}^{K} \hspace*{-1mm} \frac{ \check{p}^{(i)}_{mw} \left[ \vv{\epsilon}^{(i)}_{lq}\right]_{\kappa} \hspace*{-1mm} \mm{T}^{(i)}_l \mm{\Psi}_{lmw} \mm{T}^{(i)}_l }{ N \left(1\hspace*{-1mm}+\hspace*{-1mm}\delta^{(i)}_{lmw} \right)^2 },  \label{Eqn_T_second}
\end{align}
where $\kappa=(m-1)K+w \in \left\lbrace1,\ldots,LK\right\rbrace$ and $\vv{\epsilon}^{(i)}_{lq}$ is defined as
\begin{align}
\vv{\epsilon}^{(i)}_{lq} \triangleq \left( \mm{I}_{KL} - \mm{F}^{(i)}_{lq} \right)^{-1} \vv{f}^{(i)}_{lq}.
\end{align}
Here, the elements of matrix $\mm{F}^{(i)}_{lq}$ and vector $\vv{f}^{(i)}_{lq}$ are given by
\begin{align}
\left[ \vv{F}^{(i)}_{lq} \right]_{\kappa, \kappa^\prime} \triangleq & \frac{\check{p}^{(i)}_{mw} \check{p}^{(i)}_{mw^\prime}}{N^2\left(1+\delta_{ll w^\prime}\right)^2} \mathrm{tr} \left( \mm{\Psi}_{lmw} \mm{T}^{(i)}_l \mm{\Psi}_{lm w^\prime} \mm{T}^{(i)}_l \right), \nonumber \\
\left[ \vv{f}^{(i)}_{lq} \right]_{\kappa} \triangleq & \frac{\check{p}^{(i)}_{mw}}{N} \mathrm{tr} \left( \mm{\Psi}_{lmw} \mm{T}^{(i)}_l \mm{\Psi}_{llq} \mm{T}^{(i)}_l \right),
\end{align}
where $\kappa$ and $\kappa^\prime$ are equal to $(m-1)K+w$ and $(m-1)K+w^\prime$ with $m \in \left\lbrace 1,\ldots,L \right\rbrace$ and $w, w^\prime \in \left\lbrace 1,\ldots,K \right\rbrace$, respectively. Moreover, in (\ref{Eqn_Asy_SINR}), $\bar{ \xi}_{lq}^{(i)}$ is computed as
\begin{align}
\bar{ \xi}_{lq}^{(i)} = \frac{1}{N^2 \left( 1+\delta^{(i)}_{llq} \right)^2} \mathrm{tr} \left( \mm{\Psi}_{llq} \tilde{\mm{T}}^{(i)}_{lq} \right) \asymp \norm{\tilde{\vv{u}}_{lq}^{(i)}}^2, \label{Eqn_xi} 
\end{align}
where $\tilde{\mm{T}}^{(i)}_{lq}$ is given by (\ref{Eqn_T_second}) after replacing $\mm{\Psi}_{llq}$ with $\mm{I}_N$.
\end{theorem}
\begin{IEEEproof}
Please refer to Appendix B.
\end{IEEEproof}
%
\subsection{Asymptotic Downlink Power Allocation}
After determining the transmit powers in the dual UL, the DL power allocation is calculated using (\ref{Eqn_Vector_p}). In order to further simplify the calculation of the DL power allocation, in the following theorem, we provide deterministic expressions for the asymptotic values of the elements of vector $\vv{a}$ and matrix $\mm{A}$ in (\ref{Eqn_Vector_p}), in the large system limit, i.e., for $N \rightarrow \infty $. 
\vspace*{0mm}
\begin{theorem}\label{Theorem_Duality_Asy}
For the massive MIMO system defined in Section \ref{Sec_SysModelBench}, in the large system limit, i.e., for $N \rightarrow \infty$, the power allocation in the DL can be calculated based on the following deterministic expression
\begin{align}
\bar{\vv{p}} = \left( \mm{I}_{KL} - \mathrm{diag}\left( \bar{\vv{a}} \right) \bar{\mm{A}}^\transp \right)^{-1} \bar{\vv{a}},
\label{Eqn_Vector_p_Asy}
\end{align}
where deterministic vector $\bar{\vv{a}} \in \mathbb{C}^{LK \times 1}$ and deterministic matrix $\bar{\mm{A}} \in \mathbb{C}^{LK \times LK}$ are defined as
\begin{eqnarray}
\left[ \bar{\vv{a}} \right]_{(l-1)K+q} =  \frac{ \bar{\xi}_{lq} \left( 1+ \delta_{llq} \right)^2 {\mathrm{SINR}^\circ_{\mathrm{UL}, lq}}}{ \check{\delta}_{lq}^2 },  \label{Eqn_Asy_a}
\end{eqnarray}
\begin{eqnarray}
\left[ \bar{\mm{A}} \right]_{(l-1)K+q, (m-1)K+w} \hspace*{-1mm} = \hspace*{0mm}
\begin{cases}
  0 \hspace*{22mm} \mathrm{if} \ m=l, w = q, & \\
  \frac{ \abs{\theta_{lmq}}^2 }{ \bar{\xi}_{lq} \left( 1 + {\delta}_{llq} \right)^2 } \hspace*{8mm} \mathrm{if} \ m \neq l, w = q,  & \\
    \frac{1}{ \bar{\xi}_{lq} \left( 1 + {\delta}_{llq} \right)^2 }\bigg( \zeta_{lmw} \hspace*{0mm}+\hspace*{0mm} \frac{\abs{\theta_{lmw}}^2 \eta_{lw}}{\left(1+\delta_{llw}\right)^2} \nonumber \\
    \hspace*{0mm}-\hspace*{0mm} 2 \Re{\left\lbrace\frac{\theta_{lmw}^* \mu_{lmw} }{1+\delta_{llw}} \right\rbrace} \bigg) \hspace*{6mm} \mathrm{if} \ w \neq q.
\end{cases} 
\label{Eqn_Asy_AA}
\end{eqnarray}
\begin{algorithm}[t]\label{Algo1_MaxMin_MCARZF}
\caption{Max-Min MCA-RZF Power Allocation}
\begin{algorithmic}[1]
\STATE $\mathbf{initialization} \ \epsilon \hspace*{-.8mm}=\hspace*{-.8mm} 0.01, I_\mathrm{max}\hspace*{-1mm}=\hspace*{-1mm} 10, i \hspace*{-1mm} \leftarrow \hspace*{-1mm} 1, \ \check{p}^{(1)}_{lq} \hspace*{-1mm} \leftarrow \hspace*{-1mm} \rho_\mathrm{DL}, \forall l, q$
\WHILE{$\norm{\check{\vv{p}}^{(i+1)} - \check{\vv{p}}^{(i)}} / \norm{\check{\vv{p}}^{(i)}}   > \epsilon  $ $\mathbf{and} \ \left( i < I_\mathrm{max} \right) $}
\STATE Use (\ref{Eqn_Asy_SINR}) to compute ${\mathrm{SINR}^{{(i)}^\circ}_{\mathrm{UL}, lq}}, \forall l, q$
\STATE Substitute ${\mathrm{SINR}^{{(i)}}_{\mathrm{UL}, lq}} \leftarrow {\mathrm{SINR}^{{(i)}^\circ}_{\mathrm{UL}, lq}}$ in (\ref{MaxMin_PowerConst}) to calculate $\check{p}^{(i+1)}_{lq}, \forall l, q$
\STATE $i \leftarrow i+1$
\ENDWHILE
\end{algorithmic}
\end{algorithm}
Here, ${\mathrm{SINR}^\circ_{\mathrm{UL}, lq}}$ and $\bar{\xi}_{lq}$ are given by (\ref{Eqn_Asy_SINR}) and (\ref{Eqn_xi}), respectively, for $i=I$ with $I$ being the number of iterations performed for the calculation of the UL powers. Moreover, $\delta_{llq}=\delta^{(I)}_{llq}$ is given by (\ref{Eqn_deltak}), and $\check{\delta}_{lq}$ and $\theta_{lmw}$ are obtained as $\mathrm{tr} \left( \mm{B} \mm{T}_l \right)/N$, where $\mm{B}$ is equal to $\mm{\Psi}_{llq}$ and $\mm{R}_{lmw} \mm{\Omega}_{llw}$, respectively, $\mm{T}_l$ is given by (\ref{Eqn_T}) for $i=I$, and $\mm{\Omega}_{llw}$ is given by (\ref{Eqn_Omega}). Furthermore, $\zeta_{lmw}$, $\eta_{lw}$, and $\mu_{lmw}$ are obtained as $\mathrm{tr} \left( \mm{C} \check{\mm{T}}_{lq} \right)/N^2$ after replacing $\mm{C}$ with $\mm{R}_{lmw}$, $\check{p}_{lw} \mm{\Psi}_{llw}$, and $\check{p}_{lw} \mm{\Omega}_{lmw} \mm{R}_{llw} $, respectively, where $\check{p}_{lw} = \check{p}^{(I)}_{lw}$, and $\check{\mm{T}}_{lq}={\check{\mm{T}}}^{(I)}_{lq}$ is given by (\ref{Eqn_T_second}).
\end{theorem}
\vspace*{-0mm}
\begin{IEEEproof}
Based on (\ref{Eqn_Vector_p})-(\ref{Eqn_Mat_A}) and (\ref{Eqn_UsefulSignalAsy})-(\ref{Eqn_InterfAsy2}), expressions (\ref{Eqn_Vector_p_Asy}), (\ref{Eqn_Asy_a}) can be obtained. This completes the proof.
\end{IEEEproof}
For convenience, in Algorithm 1, we have summarized the required steps for calculating the UL Max-Min power allocation. The DL precoding vectors are calculated as $\vv{v}_{jk}= \tilde{\vv{u}}_{jk}/\norm{\tilde{\vv{u}}_{jk}}$, where $\tilde{\vv{u}}_{jk}, \forall j, k$ is given by (\ref{Eqn_Detector_PowAlloc}). Finally, (\ref{Eqn_Vector_p_Asy}) is used to compute the DL power allocation.
%
\begin{remark} As can be observed from (\ref{Eqn_Asy_SINR}), (\ref{Eqn_Vector_p_Asy}), and Algorithm 1, in contrast to the precoding vectors, which have to be updated for each channel realization cf. (\ref{Eqn_Detector_PowAlloc}), the computation of the DL power allocation requires only statistical CSI and, hence can be updated less frequently than the channel estimates. This makes the proposed scheme attractive also from the computational complexity point of view.
\end{remark}
%
\section{Numerical Results and Discussion}\label{Sec_NumResults}
\begin{figure}[tbp]
\begin{center}
\psfrag{rho}[cc][cc][0.8]{$\rho_\mathrm{DL} \ \left[\mathrm{dB}\right]$}
\includegraphics[width=0.95\linewidth, clip=true]{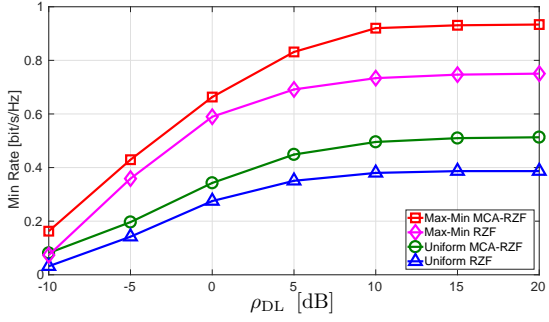}
\caption{\linespread{0.98} \small Minimum rate vs. $\rho_\mathrm{DL}=\rho_\mathrm{TR}$ for $N=60$ and $K=20$.}
\label{Fig_MinRateVsRho}
\end{center}
\end{figure}
\vspace*{0mm}
\begin{figure}[tbp]
\begin{center}
\psfrag{N}[cc][cc][0.8]{$N$}
\includegraphics[width=.95\linewidth, clip=true]{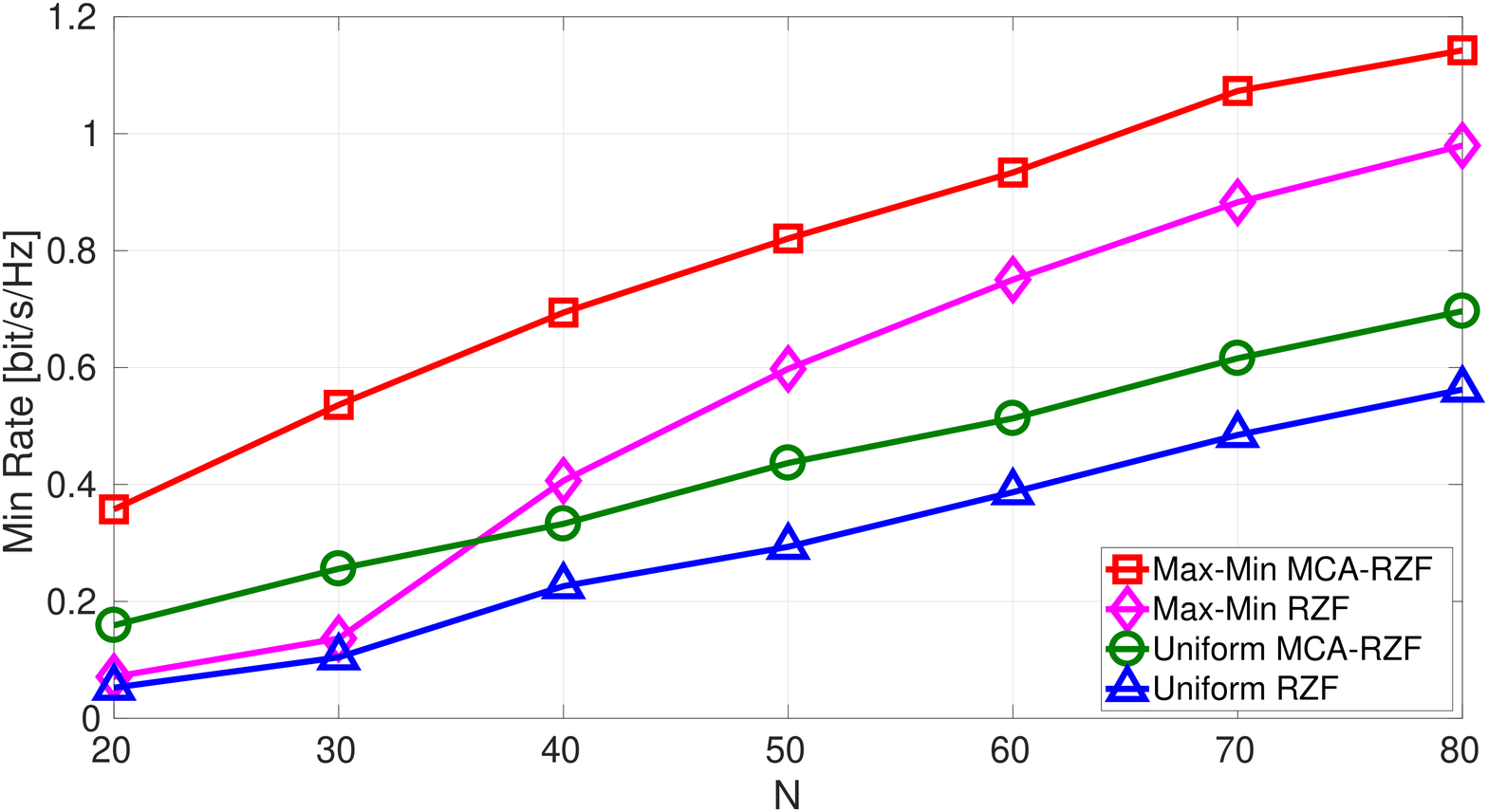}
\vspace*{0mm}
\caption{\linespread{0.98} \small Minimum rate vs. $N$ for $K=20$ and $\rho_\mathrm{DL}=\rho_\mathrm{TR}=20 \ \mathrm{dB}$.}
\label{Fig_MinRateVsN}
\end{center}
\end{figure}
\vspace*{-0mm}
In order to evaluate the performance of the proposed Max-Min MCA-RZF precoder, Monte-Carlo simulations have been performed and the results are compared to those of baseline schemes, namely the conventional RZF precoder with uniform and Max-Min power allocation as well as the MCA-RZF precoder with uniform power allocation. Here, we assume a system comprising $L\hspace*{-1mm}=\hspace*{-1mm}7$ hexagonal cells, where in each cell, there is one BS in the center and $K\hspace*{-1mm}=\hspace*{-1mm}20$ UTs which are randomly and uniformly distributed within the cell. We assume that the UTs have a maximum normalized distance of 1 to the BS and no UT is closer to the BS than $0.2$. We adopt the channel model used in \cite{Hoydis2013} which comprises large-scale fading, antenna correlation, and Rayleigh fading. Moreover, we assume that the BSs employ uniform linear arrays (ULAs) and adopt the ULA channel correlation model used in \cite{Hoydis2013}, \cite{Ngo_TCOM2013} with a normalized antenna spacing with respect to the wavelength of $\omega\hspace*{-1mm}=\hspace*{-1mm}0.5$ and number of dimensions of the antenna's physical model equal to $N$.
The considered performance metric is the network-wide minimum achievable rate given by (\ref{Eqn_MinRate_Ergo}), where the corresponding precoding vectors and transmit powers of the proposed and the baseline schemes are used. In particular, for the Max-Min RZF and Max-Min MCA-RZF precoders, the power allocation is given by (\ref{Eqn_Vector_p}) and (\ref{Eqn_Vector_p_Asy}), respectively. 

In Fig. \ref{Fig_MinRateVsRho}, the minimum achievable rate of the Max-Min MCA-RZF precoder is compared to that of the baseline precoders for $N\hspace*{-1mm}=\hspace*{-1mm}60$ as a function of $\rho_\mathrm{DL}\hspace*{-1mm}=\hspace*{-1mm}\rho_\mathrm{TR}$. As can be seen from Fig. \ref{Fig_MinRateVsRho}, the Max-Min MCA-RZF precoder achieves a substantially higher minimum rate than the MCA-RZF and the conventional RZF precoders with uniform power allocation for the considered range of $\rho_\mathrm{DL}$. For example, for $\rho_\mathrm{DL}=\rho_\mathrm{TR}=10 \ \mathrm{dB}$, the proposed Max-Min MCA-RZF precoder achieves more than twice the minimum rate of the conventional RZF precoder with uniform power allocation. Moreover, for medium-to-large values of $\rho_\mathrm{DL}$, the proposed Max-Min MCA-RZF precoder also performs considerably better than the Max-Min RZF precoder. 

In Fig. \ref{Fig_MinRateVsN}, the minimum achievable rate of the investigated precoders as a function of $\hspace*{-.5mm} N \hspace*{-.5mm}$ is depicted for $\rho_\mathrm{DL}$ $=$ $\rho_\mathrm{TR}$ $=$ $20$ $ \mathrm{dB}$. As can be observed, the proposed Max-Min MCA-RZF precoder achieves a considerably higher minimum rate than the baseline precoders for the entire considered range of $N$, which underlines the importance of both a proper power allocation and multi-cell aware precoding for achieving a high performance.
%
\appendices
%
\section*{Appendix A - Proof of Theorem 1} \label{Theo1Proof}
First, we define the estimation error of the channel between the $k$th UT in the $l$th cell and the $j$th BS as $\tilde{\vv{g}}_{jlk} = {\vv{g}}_{jlk} - \hat{\vv{g}}_{jlk}$ with its covariance matrix given by
\begin{align}
\mm{\Delta}_{jlk} \triangleq & \mathbb{E}_{\mm{G}} \left\lbrace \tilde{\vv{g}}_{jlk} \tilde{\vv{g}}_{jlk}^\herm \right\rbrace = \mathbb{E}_{\mm{G}} \left\lbrace {\mm{g}}_{jlk} {\mm{g}}_{jlk}^\herm \right\rbrace - \mathbb{E}_{\mm{G}} \left\lbrace \hat{\mm{g}}_{jlk} \hat{\mm{g}}_{jlk}^\herm \right\rbrace \nonumber \\
& = \mm{R}_{jlk} - \mm{\Psi}_{jlk},
\label{Eqn_Delta_jl}
\end{align}
where we exploited the uncorrelatedness of $\hat{\vv{g}}_{jlk}$ and $\tilde{\vv{g}}_{jlk}$ for LMMSE estimation \cite{Kay_StSiProDet2013}, and $\mm{\Psi}_{jlk}=\mathbb{E}_{\mm{G}} \left\lbrace \hat{\mm{g}}_{jlk} \hat{\mm{g}}_{jlk}^\herm \right\rbrace$ is given by (\ref{Eqn_Delta}) for $q=k$, where we used (\ref{Eqn_g_hat}) and \cite[Theorem 7]{Evans2000}. Next, we substitute ${\vv{g}}_{jlk} = \hat{\vv{g}}_{jlk} + \tilde{\vv{g}}_{jlk}$ into (\ref{Eqn_SINR_UL1}), apply (\ref{Eqn_Delta_jl}), exploit $\mathbb{E}_{\mm{G}}\left\lbrace x \right\rbrace = \mathbb{E}_{\hat{\mm{G}}}\left\lbrace \mathbb{E}_{\tilde{\mm{G}}}\left\lbrace x \big| \hat{\mm{G}} \right\rbrace \right\rbrace$, and obtain the following approximation for the SINR in the UL:
\begin{align}
& \mathrm{SINR}^\mathrm{UL}_{jk} \approx \nonumber \\
&  \frac{ \check{p}_{jk} \left| \mathbb{E}_{\hat{\mm{G}}} \left\lbrace \mm{v}_{jk}^\herm \hat{\vv{g}}_{jjk} \right\rbrace \right|^2 }{ \displaystyle \mathop{ \sum_{l=1}^{L} \sum_{q=1}^{K}}_{(l, q) \neq (j, k)} \mathbb{E}_{\hat{\mm{G}}} \left\lbrace \mm{v}_{jk}^\herm \left( \check{p}_{lq} \hat{\vv{g}}_{jlq} \hat{\vv{g}}_{jlq}^\herm + \check{p}_{lq} \mm{\Delta}_{jlq}+ \mm{I}_N \right) \mm{v}_{jk} \right\rbrace },
\label{Eqn_SINR_UL_2}
\end{align}
where we took $\mm{v}_{jk}^\herm \mm{v}_{jk} = 1$ into account, and neglected the term $\check{p}_{jk} \mathbb{E}_{\mm{G}} \left\lbrace \big| {\mm{v}_{jk}^\herm \vv{g}_{jjk}} \big| ^2 \right\rbrace - \check{p}_{jk} \Big| \mathbb{E}_{\mm{G}} \left\lbrace {\mm{v}_{jk}^\herm \vv{g}_{jjk} } \right\rbrace \Big| ^2 $ in the denominator, since it vanishes according to \cite{Hoydis2013} in the large system limit. Here, $\hat{\mm{G}}=\left[ \hat{\mm{G}}_1 \cdots \hat{\mm{G}}_L\right]$ and $\tilde{\mm{G}}=\left[ \tilde{\mm{G}}_1 \cdots \tilde{\mm{G}}_L\right]$, where $\hat{\mm{G}}_l=\left[\hat{\vv{g}}_{l11} \ldots \hat{\vv{g}}_{lLK} \right]$ and $\tilde{\mm{G}}_l=\left[\tilde{\vv{g}}_{l11} \ldots \tilde{\vv{g}}_{lLK} \right]$. In the large system limit, i.e., for $N \rightarrow \infty$, the operands of the expectation operators in the nominator and denominator of (\ref{Eqn_SINR_UL_2}) become deterministic and $\mathrm{SINR}^\mathrm{UL}_{jk}$ becomes a generalized Rayleigh quotient, which is maximized by $\mm{v}_{jk}= \tilde{\mm{u}}_{jk}/ \norm{ \tilde{\mm{u}}_{jk} }$, where $\tilde{\mm{u}}_{jk}$ is given by (\ref{Eqn_Detector_PowAlloc}). In particular, it can be observed from (\ref{Eqn_SINR_UL_2}) that for given transmit powers, the SINR of each UT in the UL can be \emph{individually} maximized by its corresponding optimal detector vector. Considering this and taking into account that the log function is monotonically increasing, we conclude that $\tilde{\mm{u}}_{jk} / \norm{ \tilde{\mm{u}}_{jk} }$ is the solution to (\ref{Eqn_MaxMinOptProblem_UL}) for given transmit powers. This completes the proof.
%
\vspace*{-2mm}
\section*{Appendix B - Proof of Theorem 2} \label{Theo2Proof}
\vspace*{-2mm}
First, we derive a deterministic expression for $ \xi^{(i)}_{lq} = \norm{ \tilde{\vv{u}}^{(i)}_{lq} }^2$ for $N\rightarrow \infty$
\begin{align}
& \frac{1}{N^2} \hat{\vv{g}}_{llq}^\herm (\mm{\Lambda}_{l}^{(i)})^{-2} \hat{\vv{g}}_{llq} = \frac{\hat{\vv{g}}_{llq}^\herm (\mm{\Lambda}^{(i)}_{lq})^{-2} \hat{\vv{g}}_{llq}}{N^2 \left( 1+ \frac{1}{N} \hat{\vv{g}}_{llq}^\herm (\mm{\Lambda}^{(i)}_{lq})^{-1} \hat{\vv{g}}_{llq} \right)^2} \nonumber \\
& \asymp \frac{\mathrm{tr}\left( \mm{\Psi}_{llq} {\tilde{\mm{T}}}^{(i)}_{lq} \right)}{N^2\left(1+\delta^{(i)}_{llq}\right)^2}= \bar{\xi}^{(i)}_{lq} ,
\label{Eqn_UsefulSignalAsy}
\end{align}
where $\mm{\Psi}_{llq}$ is given by (\ref{Eqn_Delta}) for $j=l$, and $\delta^{(i)}_{llq}$ and ${\tilde{\mm{T}}}^{(i)}_{lq}$ are given by (\ref{Eqn_deltak}) for $m=l$ and $w=q$, and (\ref{Eqn_T_second}), respectively, and we used the matrix inversion lemma \cite{Horn2013}, the rank-1 perturbation lemma \cite[Lemma 14.3]{Couillet_Book2011}, \cite[Theorem 7]{Evans2000}, \cite[Theorem 1]{Wagner2012}, and \cite[Theorem 2]{Hoydis2013}. In (\ref{Eqn_UsefulSignalAsy}), $\mm{\Lambda}^{(i)}_{l}$ is defined as
\begin{align}
\mm{\Lambda}^{(i)}_l \triangleq \hspace*{-0.5mm}\frac{1}{N} \hspace*{-0.5mm} \sum_{m=1}^L \hspace*{-.5mm} \sum_{w=1}^{K} \check{p}^{(i)}_{mw} \left(  \hat{\vv{g}}_{lmw} \hat{\vv{g}}_{lmw}^\herm + \mm{\Delta}_{lmw} \hspace*{-0.5mm} \right) \hspace*{-0.5mm}+\hspace*{-0.5mm} \frac{1}{N } \mm{I}_N \hspace*{-0.5mm},
\end{align}
and $\mm{\Lambda}^{(i)}_{lq}$ is given by $\mm{\Lambda}^{(i)}_{lq} = \mm{\Lambda}^{(i)}_l - \check{p}^{(i)}_{lq} \hat{\vv{g}}_{llq} \hat{\vv{g}}^\herm_{llq}/N$.
Next, we derive a deterministic equivalent for the useful signal power of the $q$th UT in the $l$th cell in the UL. According to (\ref{Eqn_SINR_UL1}), and taking into account that $\vv{v}^{(i)}_{lq} = \tilde{\vv{u}}^{(i)}_{lq} / \norm{\tilde{\vv{u}}^{(i)}_{lq}}$, we have
\begin{align}
& (\vv{v}^{(i)}_{lq})^\herm \vv{g}_{llq} = \frac{ 1 }{N \sqrt{ \xi^{(i)}_{lq} } } \hat{\vv{g}}_{llq}^\herm (\mm{\Lambda}^{(i)}_{l})^{-1} \vv{g}_{llq} \hspace*{-1mm} \nonumber \\
& = \frac{ \frac{1}{N} \hat{\vv{g}}_{llq}^\herm (\mm{\Lambda}^{(i)}_{lq})^{-1} \vv{g}_{llq} }{ \sqrt{ \xi^{(i)}_{lq} } \hspace*{-0.5mm} \left( 1 \hspace*{-1mm}+\hspace*{-1mm} \frac{1}{N} \hat{\vv{g}}_{llq}^\herm (\mm{\Lambda}^{(i)}_{lq})^{-1} \hat{\vv{g}}_{llq} \right) } \hspace*{-0.5mm} \asymp \hspace*{-0.5mm} \frac{ \check{\delta}^{(i)}_{lq} }{ \sqrt{ \bar{ \xi}^{(i)}_{lq} } \hspace*{-0.5mm} \left(1\hspace*{-1mm}+\hspace*{-1mm}\delta^{(i)}_{llq}\right)}, 
\label{Eqn_UsefulAsy}
\end{align}
where $\check{\delta}^{(i)}_{lq}=\mathrm{tr} \left( \mm{\Psi}_{llq} \mm{T}^{(i)}_l \right) / N$ with $\mm{\Psi}_{llq}$ given by (\ref{Eqn_Delta}) for $j=l$, and $\delta^{(i)}_{llq}$ is given by (\ref{Eqn_deltak}) for $m=l$ and $w=q$, and we used the matrix inversion lemma \cite{Horn2013}, the rank-1 perturbation lemma \cite[Lemma 14.3]{Couillet_Book2011}, \cite[Theorem 7]{Evans2000}, and \cite[Theorem 1]{Wagner2012}.
Now, we derive the asymptotic expression for the interference power of the $q$th UT in the $l$th cell in the UL for $N \rightarrow \infty$. According to (\ref{Eqn_SINR_UL1}), the interference power of the $q$th UT in the $l$th cell in the UL is given by 
\begin{align}
& \mathop{\sum_{m=1}^{L}\sum_{w=1}^{K}}_{ (m, w) \neq (l, q)}  \check{p}^{(i)}_{mw} \left| (\vv{v}^{(i)}_{lq})^\herm \vv{g}_{lmw} \right|^2 = \hspace*{-3mm} \sum_{m=1, m \neq l}^{L} \check{p}^{(i)}_{mq} \left| (\vv{v}^{(i)}_{lq})^\herm \vv{g}_{lmq} \right|^2 \nonumber \\
& + \sum_{m=1}^{L} \sum_{w=1, w \neq q}^K \check{p}^{(i)}_{mw} \left| (\vv{v}^{(i)}_{lq})^\herm \vv{g}_{lmw} \right|^2.
\label{Eqn_InterfAsy1}
\end{align}
Now, taking into account that $\vv{v}^{(i)}_{lq} = \tilde{\vv{u}}^{(i)}_{lq} / \norm{ \tilde{\vv{u}}^{(i)}_{lq} }$, we have the following expression for the first term on the right hand side of (\ref{Eqn_InterfAsy1})
\begin{align}
& \sum_{m=1, m \neq l}^{L} \hspace*{-3mm} \check{p}^{(i)}_{mq} \left| (\vv{v}^{(i)}_{lq})^\herm \vv{g}_{lmq} \right|^2 \hspace*{-2mm}=\hspace*{-1mm} \frac{ \check{p}^{(i)}_{mq} \left| \hat{\vv{g}}_{llq}^\herm (\mm{\Lambda}^{(i)}_{lq})^{-1} \vv{g}_{lmq} \right|^2 }{ \hspace*{-.5mm} \xi^{(i)}_{lq} \hspace*{-.5mm} N^2 \hspace*{-.5mm} \left( \hspace*{-.5mm} 1 \hspace*{-1mm}+\hspace*{-1mm} \frac{1}{N} \hat{\vv{g}}_{llq}^\herm (\mm{\Lambda}^{(i)}_{lq})^{-1} \hat{\vv{g}}_{llq} \right)^2 } \nonumber \\
& \asymp \frac{ \check{p}^{(i)}_{mq} \left| \theta^{(i)}_{lmq} \right|^2 }{ \bar{ \xi}^{(i)}_{lq} \left(1+\delta^{(i)}_{llq} \right)^2},
\label{Eqn_InterfAsy}
\end{align}
where $ \theta^{(i)}_{lmq} = \mathrm{tr} \left( \mm{R}_{lmq} \mm{\Omega}_{llq} \mm{T}^{(i)}_l \right)/N$, and we used the matrix inversion lemma \cite{Horn2013}, the rank-1 perturbation lemma \cite[Lemma 14.3]{Couillet_Book2011}, \cite[Theorem 7]{Evans2000}, and \cite[Theorem 1]{Wagner2012}. Similarly, the second term on the right hand side of (\ref{Eqn_InterfAsy1}) can be rewritten as
\begin{align}
& \check{p}^{(i)}_{mw} \hspace*{-.5mm} \left| (\vv{v}^{(i)}_{lq})^\herm \vv{g}_{lmw} \right|^2 \hspace*{-2.5mm} = \hspace*{-.5mm} \frac{ \check{p}^{(i)}_{mw} \hat{\vv{g}}_{llq}^\herm (\mm{\Lambda}_{lq}^{(i)})^{-1} \vv{g}_{lmw} \vv{g}_{lmw}^\herm (\mm{\Lambda}^{(i)}_{lq})^{-1} \hat{\vv{g}}_{llq}}{ \hspace*{-.5mm} \xi^{(i)}_{lq} N^2  \hspace*{-.5mm} \left( \hspace*{-.5mm} 1 \hspace*{-.5mm} + \hspace*{-.5mm} \frac{1}{N} \hat{\vv{g}}_{llq}^\herm (\mm{\Lambda}^{(i)}_{lq})^{-1} \hat{\vv{g}}_{llq} \hspace*{-.5mm} \hspace*{-.5mm} \right)^2} \nonumber \\
& \asymp \frac{ \check{p}^{(i)}_{mw} {\vv{g}}_{lmw}^\herm (\mm{\Lambda}^{(i)}_{lq})^{-1} \mm{\Psi}_{llq} (\mm{\Lambda}^{(i)}_{lq})^{-1} \vv{g}_{lmw}}{ \bar{ \xi}^{(i)}_{lq} N^2 \left( 1  + \delta^{(i)}_{llq} \right)^2} \nonumber \\
& = \frac{\check{p}^{(i)}_{mw}}{ \bar{ \xi}^{(i)}_{lq}   \left(  1 + \delta^{(i)}_{llq}  \right)^2} \Bigg( \hspace*{-.5mm} \frac{ \vv{g}_{lmw}^\herm (\mm{\Lambda}^{(i)}_{lqw})^{-1} \mm{\Psi}_{llq} (\mm{\Lambda}^{(i)}_{lqw})^{-1} \vv{g}_{lmw} }{N^2} \nonumber \\
& + \frac{\left|\vv{g}_{lmw}^\herm (\mm{\Lambda}^{(i)}_{lqw})^{-1} \hat{\vv{g}}_{llw}\right|^2 \hat{\vv{g}}_{llw}^\herm (\mm{\Lambda}^{(i)}_{lqw})^{-1} \mm{\Psi}_{llq} (\mm{\Lambda}^{(i)}_{lqw})^{-1} \hat{\vv{g}}_{llw} }{N^4 \left(1+\delta^{(i)}_{llw}\right)^2} \nonumber \\
& - 2 \mathfrak{R} \bigg\{ \hat{\vv{g}}_{llw}^\herm (\mm{\Lambda}^{(i)}_{lqw})^{-1} \vv{g}_{lmw} \vv{g}_{lmw}^\herm (\mm{\Lambda}^{(i)}_{lqw})^{-1}  \mm{\Psi}_{llq} (\mm{\Lambda}^{(i)}_{lqw})^{-1} \hat{\vv{g}}_{llw} \nonumber \\
& \bigg/ \left( N^3 \hspace*{-.5mm} \left( \hspace*{-.5mm} 1+\delta^{(i)}_{llw} \hspace*{-.5mm} \right) \right) \hspace*{-.5mm} \bigg\} \hspace*{-.5mm} \Bigg) \hspace*{-.5mm} \asymp  \nonumber \\
& \hspace*{-.5mm} \frac{\check{p}^{(i)}_{mw}}{ \bar{ \xi}^{(i)}_{lq} \left( \hspace*{-.5mm} 1 \hspace*{-.5mm}+\hspace*{-.5mm}\delta^{(i)}_{llq}  \right)^2} \hspace*{-.5mm} \Bigg( \hspace*{-.5mm} \zeta^{(i)}_{lmw} \hspace*{-1mm} + \hspace*{-.5mm} \frac{\left| \theta^{(i)}_{lmw} \right|^2 \hspace*{-.5mm} \eta^{(i)}_{lw} }{ \hspace*{-.5mm} \left( \hspace*{-.5mm} 1\hspace*{-.5mm}+\hspace*{-.5mm}\delta^{(i)}_{llw} \hspace*{-.5mm} \right)^2} \hspace*{-.5mm} - \hspace*{-.5mm} 2 \mathfrak{R} \hspace*{-.5mm} \left\lbrace \hspace*{-.5mm} \frac{(\theta^{(i)}_{lmw})^* \mu^{(i)}_{lmw} }{1 \hspace*{-.5mm} + \hspace*{-.5mm} \delta^{(i)}_{llw}} \hspace*{-.5mm} \right\rbrace \hspace*{-.5mm} \Bigg)
\label{Eqn_InterfAsy2}
\end{align}
where $\delta_{llq}$ is given by (\ref{Eqn_deltak}) for $m=l$ and $w=q$, and $\zeta^{(i)}_{lmw}$, $\eta^{(i)}_{lw}$, and $\mu^{(i)}_{lmw}$ are obtained as $\mathrm{tr} \left( \mm{C} \check{\mm{T}}^{(i)}_{lq} \right)/N^2$ after replacing $\mm{C}$ with $\mm{R}_{lmw}$, $\check{p}^{(i)}_{lw} \mm{\Psi}_{llw}$, and $\check{p}^{(i)}_{lw} \mm{\Omega}_{lmw} \mm{R}_{llw} $, respectively. Here, we used again the matrix inversion lemma \cite{Horn2013}, the rank-1 perturbation lemma \cite[Lemma 14.3]{Couillet_Book2011}, \cite[Theorem 7]{Evans2000}, \cite[Theorem 1]{Wagner2012}, and \cite[Theorem 2]{Hoydis2013}. Taking the dominated convergence theorem \cite{Billingsley_ProbMeasureBook1995} and continuous mapping theorem \cite{VanDerVaart_AsyStatBook2000} into account and considering (\ref{Eqn_SINR_UL1}) and (\ref{Eqn_UsefulSignalAsy})-(\ref{Eqn_InterfAsy2}) we obtain (\ref{Eqn_Asy_SINR}). This completes the proof.
%

%
\ifCLASSOPTIONcaptionsoff
  \newpage
\fi
\bibliographystyle{IEEEtran}
\bibliography{Massive_MIMO}
\end{document}